\documentclass[12pt]{article}
\usepackage{amsmath}
\usepackage{amsfonts}
\usepackage{amssymb}
\usepackage{latexsym}
\usepackage{graphicx}
\usepackage[english]{babel}
\input epsf.sty

\def\tr{\mbox{Tr}}
\def\str{\mbox{STr}}

\begin{document}
{}~ {}~ \hfill\vbox{\hbox{USTC-ICTS-08-07}}\break

\vskip .4cm \centerline{\Large \bf M2-branes Coupled to Antisymmetric Fluxes} \vskip .6cm
\medskip

\vspace*{4.0ex}
\centerline{\large Miao Li\footnote{E-mail: mli@itp.ac.cn}, Tower Wang\footnote{E-mail: wangtao218@itp.ac.cn}}
\vspace*{4.0ex}

\centerline{\it Interdisciplinary Center for Theoretical Study,}
\centerline{\it University of Science and Technology of China,}
\centerline{\it Hefei, Anhui 230026, China}
\centerline{\it  Institute of Theoretical Physics, Chinese Academy of Sciences,}
\centerline{\it P. O. Box 2735£¬ Beijing 100080, China}

\vspace*{5.0ex} \centerline{\bf Abstract}\bigskip

By turning on antisymmetric background fluxes, we study how multiple
M2-branes are coupled to them.  Our investigation concentrates on the
gauge invariance conditions for
the Myers-Chern-Simons action. Furthermore, the dimensional
reduction of M2-branes to D2-branes introduces more constraints on the
newly introduced tensors. Particularly, for the theory based on
$\mathcal{A}_4$ algebra, we are able to fix all components of them
up to an overall normalization constant. These results can not be simply obtained
from the previously proposed cubic matrix representation for this
algebra. We also comment on cubic matrices as the representations of
3-algebras.

\vfill \eject

\baselineskip=18pt

\section{Introduction}
The dynamics of D-branes was well studied a decade ago
\cite{Polchinski:1995mt,Li:1995pq,Douglas:1995bn,Tseytlin:1997csa,Myers:1999ps},
which has driven huge progress of string theory. In contrast,
multiple M-branes are more untamable. Recently, a 3-dimensional
field theory for multiple M2-branes was proposed by Bagger and
Lambert \cite{Bagger:2006sk,Bagger:2007jr,Bagger:2007vi} and
Gustavsson \cite{Gustavsson:2007vu,Gustavsson:2008dy}. To check this
theory, it is important to do the dimensional reduction from
M2-branes to D2-branes. The reduction has been investigated from
different viewpoints in
\cite{Gustavsson:2007vu,Mukhi:2008ux,Gran:2008vi,Ho:2008ei}. By
virtue of the Nambu-Poisson algebra, Ho and Matsuo et.al. have
carefully worked over the relations between multiple M2-brans and a
M5-brane \cite{Ho:2008bn,Ho:2008nn,Ho:2008ve,Ho:2008ei}.

The construction of Bagger-Lambert-Gustavsson model relies on a
3-algebra with a positive-definite metric, namely the
$\mathcal{A}_4$ algebra. It was conjectured by \cite{Ho:2008bn} and
confirmed by \cite{Papadopoulos:2008sk,Gauntlett:2008uf} that the
BLG theory is unique \cite{Gustavsson:2008dy,Bandres:2008vf},
because all finite dimensional 3-algebras with a positive-definite
metric are direct sums of $\mathcal{A}_4$ with trivial algebras
\cite{Ho:2008bn,Gauntlett:2008uf}. However, if we do not require the
metric to be positive-definite, there are still a rich class of
models \cite{Awata:1999dz}. In particular, very recently, a class of
models based on 3-algebras with a Lorentzian metric\footnote{The
constructions in \cite{Gomis:2008uv,Benvenuti:2008bt,Ho:2008ei} are
more general, which lead to a 3-algebra with a Lorentzian metric if
one starts with a compact semi-simple Lie 2-algebra as a special
case.} were studied by several groups
\cite{Gomis:2008uv,Benvenuti:2008bt,Ho:2008ei}. Besides considering
Lorentzian metrics, various attempts to extending the BLG model were
also made during the past few months
\cite{Gran:2008vi,Ho:2008bn,Morozov:2008cb,Morozov:2008rc}.

In \cite{Lambert:2008et,Distler:2008mk,Fuji:2008yj}, there are some
interesting discussions relating BLG theory to M2-branes on an
obifold (or namely an ``M-fold'' \cite{Distler:2008mk}). As a
partial list, other aspects of BLG theory were extensively studied
in
\cite{Gomis:2008cv,Bergshoeff:2008cz,Hosomichi:2008qk,Papadopoulos:2008gh,Honma:2008un,Krishnan:2008zm,Song:2008bi,Jeon:2008bx}.

Our interest in this paper is to consider the effects of the
background antisymmetric tensor fields on multiple M2-branes. It is
well-known that in the world-volume theory of D-branes, when the
background fluxes are switched on, there will be additional
Chern-Simons terms \cite{Li:1995pq,Douglas:1995bn}, through which
the background fields couple not only to the internal gauge fields
but also to the world-volume scalar fields \cite{Myers:1999ps}. In
string theory, the antisymmetric tensor fields are sourced by
various dimensional D-branes or strings. In M-theory, these are
3-forms $C_{(3)}$ and 6-forms $C_{(6)}$, they are dual to each other
and are coupled to M2-branes and M5-branes respectively. The
Chern-Simons terms we want to discuss describe interactions between
world-volume fields and $C_{(3)}$, $C_{(6)}$. To distinguish them
from those Chern-Simons terms purely of internal gauge fields (that
is, the terms appearing in Bagger and Lambert's papers
\cite{Bagger:2007jr,Bagger:2007vi}), we will call them
Myers-Chern-Simons terms. Since M2-branes can be obtained by lifting
D2-branes in string theory, it is important to understand these
terms for multiple M2-branes.

First of all, we suggest that to lowest order, the
Myers-Chern-Simons action for multiple M2-branes takes the form
\begin{eqnarray}\label{MCS0}
\nonumber \mathcal{S}_{MCS}&=&\int d^3\sigma\left[\lambda_{1}\epsilon^{\lambda\mu\nu}C_{IJK}\str(T^{a}T^{b}T^{c})D_{\lambda}X^{I}_{a}D_{\mu}X^{J}_{b}D_{\nu}X^{K}_{c}\right.\\
\nonumber &&\left.+\lambda_{2}\epsilon^{\lambda\mu\nu}C_{IJKLMN}\str([T^{d},T^{e},T^{f}]T^{a}T^{b}T^{c})X^{I}_{d}X^{J}_{e}X^{K}_{f}D_{\lambda}X^{L}_{a}D_{\mu}X^{M}_{b}D_{\nu}X^{N}_{c}\right].\\
\end{eqnarray}
When writing down this action, we only take the lowest order terms
into consideration, although it is expectable that a full
Myers-Chern-Simons action will involve higher order terms of
$C_{(3)}$, $C_{(6)}$ and $A_{\lambda ab}$ and their derivatives.
Here $\epsilon^{\lambda\mu\nu}$ ($\lambda,\mu,\nu=1,2,3$) is the
Levi-Civita symbol with $\epsilon^{123}=1$ along the world-volume
directions. The coefficients $\lambda_{1}$ and $\lambda_{2}$ depend
on conventions, so we leave them as unfixed parameters at present.
The covariant derivative with respect to the internal gauge field
$A_{\lambda ab}$ is \cite{Bagger:2007jr,Bagger:2007vi}
\begin{equation}\label{covderiv}
D_{\lambda}X^{I}_{a}=\partial_{\lambda}X^{I}_{a}-A_{\lambda dc}f^{dcb}_{~~~~a}X^{I}_{b}.
\end{equation}
In this paper, we always use ``$\str$'' to denote the symmetrized
trace as they did in \cite{Tseytlin:1997csa,Myers:1999ps}, e.g. (not
one appearing in (\ref{MCS0}))
\begin{eqnarray}
\nonumber &&\str([T^{a},T^{b},T^{c}]T^{d}T^{e})\\
\nonumber &=&\frac{1}{3!}\left[\tr([T^{a},T^{b},T^{c}]T^{d}T^{e})+\tr(T^{d}T^{e}[T^{a},T^{b},T^{c}])+\tr(T^{e}[T^{a},T^{b},T^{c}]T^{d})\right.\\
&&\left.+\tr([T^{a},T^{b},T^{c}]T^{e}T^{d})+\tr(T^{e}T^{d}[T^{a},T^{b},T^{c}])+\tr(T^{d}[T^{a},T^{b},T^{c}]T^{e})\right],
\end{eqnarray}
and ``\tr'' to denote an ordinary trace. But, in the present case of
multiple M2-branes, we do not know how to perform such a trace
``\tr'' unless one gives the correct representation of all
generators \{$T^{a}$\}. To circumambulate this difficulty, let us
introduce some succinct notations
\begin{eqnarray}\label{symtrace}
\nonumber g^{abc}&=&g^{(abc)}=\str(T^{a}T^{b}T^{c}),\\
d^{abcd}&=&d^{(abcd)}=\str(T^{a}T^{b}T^{c}T^{d}).
\end{eqnarray}
The tensors $g^{abc}$ and $d^{abcd}$ will be very useful. They are
completely symmetric under the permutation of indices as indicated.
In our conventions, indices inside round brackets $(a_1,...,a_n)$
will always be understood as symmetrized with unit weight, i.e. with
a factor $1/n!$. We will use the same unit weight convention to
anti-symmetrize indices inside square brackets $[a_1, ... , a_n]$.

Then in terms of notations (\ref{symtrace}), the lowest order
Myers-Chern-Simons action becomes
\begin{eqnarray}\label{MCS}
\nonumber \mathcal{S}_{MCS}&=&\int d^3\sigma\left[\lambda_{1}\epsilon^{\lambda\mu\nu}C_{IJK}g^{abc}D_{\lambda}X^{I}_{a}D_{\mu}X^{J}_{b}D_{\nu}X^{K}_{c}\right.\\
&&\left.+\lambda_{2}\epsilon^{\lambda\mu\nu}C_{IJKLMN}d^{gabc}f^{def}_{~~~~g}X^{I}_{d}X^{J}_{e}X^{K}_{f}D_{\lambda}X^{L}_{a}D_{\mu}X^{M}_{b}D_{\nu}X^{N}_{c}\right].
\end{eqnarray}
This form of action may be still correct when background fields
$C_{(3)}$, $C_{(6)}$ are functionals of non-Abelian scalars
$X^{I},X^{J}$, etc. But in our following investigation we will
consider constant background fields for simplicity. Furthermore, to
systematically neglect other terms induced by the metric of
spacetime, we work in a flat spacetime background.

The remaining of this paper is organized as follows. In section
\ref{sectGI} we derive the general conditions of gauge invariance
for the above Myers-Chern-Simons action. After a brief review of Lie
2-algebras in section \ref{lie2}, we solve the gauge invariance
conditions and the constraints arising from dimensional reduction,
hence fix almost all of the components appearing in
(\ref{symtrace}). This is done in concrete examples, i.e., section
\ref{Euclidean} for $\mathcal{A}_4$, and section \ref{Lorentzian}
for 3-algebras with a Lorentzian metric. Failing to reobtain the
above results in terms of cubic matrices, in section
\ref{cubicmatrix}, we present some tentative thoughts on cubic
matrices as representations of 3-algebras. Section \ref{conclusion}
is a short conclusion.

\section{Gauge Invariance}\label{sectGI}
One of the most apparent constraints on action (\ref{MCS}) comes
from the requirement that it should be gauge invariant. Under the
gauge transformation
\begin{equation}
\delta X^{I}_{a}=\Lambda_{fe}f^{fed}_{~~~~a}X^{I}_{d},~~~~\delta(D_{\lambda}X^{I}_{a})=\Lambda_{fe}f^{fed}_{~~~~a}D_{\lambda}X^{I}_{d},
\end{equation}
the first part of (\ref{MCS}) changes as
\begin{eqnarray}\label{ginpart1}
\nonumber &&\delta\left[\epsilon^{\lambda\mu\nu}C_{IJK}g^{abc}D_{\lambda}X^{I}_{a}D_{\mu}X^{J}_{b}D_{\nu}X^{K}_{c}\right]\\
&=&\epsilon^{\lambda\mu\nu}C_{IJK}\Lambda_{fe}\left(g^{dbc}f^{fea}_{~~~~d}+g^{dac}f^{feb}_{~~~~d}+g^{dab}f^{fec}_{~~~~d}\right)D_{\lambda}X^{I}_{a}D_{\mu}X^{J}_{b}D_{\nu}X^{K}_{c},
\end{eqnarray}
while the gauge transformation of the second part gives
\begin{eqnarray}\label{ginpart2}
\nonumber &&\delta\left[\epsilon^{\lambda\mu\nu}C_{IJKLMN}d^{gabc}f^{def}_{~~~~g}X^{I}_{d}X^{J}_{e}X^{K}_{f}D_{\lambda}X^{L}_{a}D_{\mu}X^{M}_{b}D_{\nu}X^{N}_{c}\right]\\
\nonumber &=&\epsilon^{\lambda\mu\nu}C_{IJKLMN}\Lambda_{ji}\left[d^{gabc}(f^{jid}_{~~~~h}f^{efh}_{~~~~g}-f^{jie}_{~~~~h}f^{dfh}_{~~~~g}+f^{jif}_{~~~~h}f^{deh}_{~~~~g})\right.\\
\nonumber &&\left.+(d^{ghbc}f^{jia}_{~~~~h}+d^{ghac}f^{jib}_{~~~~h}+d^{ghab}f^{jic}_{~~~~h})f^{def}_{~~~~g}\right]X^{I}_{d}X^{J}_{e}X^{K}_{f}D_{\lambda}X^{L}_{a}D_{\mu}X^{M}_{b}D_{\nu}X^{N}_{c}.\\
\end{eqnarray}
Here we only consider constant background fields. In general
$C_{(3)}$, $C_{(6)}$ are functionals of non-Abelian scalars
$X^{I},X^{J}$, etc, then more terms will be involved in
(\ref{ginpart1}) and (\ref{ginpart2}).

According to (\ref{ginpart1}) and (\ref{ginpart2}), the gauge
invariance of the Myers-Chern-Simons terms (\ref{MCS}) imposes the
conditions
\begin{eqnarray}\label{gincond}
\nonumber g^{d(ab}f^{c)fe}_{~~~~d}&=&0,\\
d^{gabc}f^{ji[d}_{~~~~h}f^{ef]h}_{~~~~g}+d^{gh(ab}f^{c)ji}_{~~~~h}f^{def}_{~~~~g}&=&0.
\end{eqnarray}
This system of equations appears to be over-determined because there
are much more equations than independent unknowns. However, these
equations are not independent of each other. Obviously they are
trivially satisfied for
\begin{equation}\label{trivialsol}
g^{abc}=0,~~~~d^{abcd}=0.
\end{equation}

Of course this is not the solution we want. We shall see in all
concrete examples, the gauge invariance conditions (\ref{gincond})
allow for non-trivial solutions. With the help of computer, given
the structure constants $f^{abc}_{~~~~d}$ of an algebra, we can
solve numerous linear equations (\ref{gincond}) by brute-force, then
get the simplified conditions for $g^{abc}$ and $d^{abcd}$. This is
exactly what we will do in sections \ref{Euclidean} and
\ref{Lorentzian}. Before doing that, to establish  our conventions
and notations, we briefly collect some well-known facts about Lie
2-algebras in section \ref{lie2}.

\section{Lie 2-Algebras}\label{lie2}
In string theory, a $U(N)$ internal gauge symmetry is present on $N$
coincident D-branes, reducing to $SU(N)$ when we do not consider the
center-of-mass (or zero) mode. As concrete examples, we start with
$U(2)$ and $U(3)$.

\subsection{$SU(2)$ and $U(2)$}
The $SU(2)$ algebra is well-known,
\begin{equation}\label{SUalgebra}
[T^{a},T^{b}]=f^{ab}_{~~~c}T^{c},
\end{equation}
where $a,b,c=1,2,3$, and the structure constants
$f^{ab}_{~~~c}=f\epsilon^{abc}$ with $\epsilon^{123}=1$. The
representation (planar) matrices of its generators are proportional
to the Pauli matrices,
\begin{equation}\label{Pauli}
T^{1}=\frac{f}{2}\left(
                   \begin{array}{cc}
                     0 & i\\
                     i & 0 \\
                   \end{array}
                 \right),
~~~~T^{2}=\frac{f}{2}\left(
                   \begin{array}{cc}
                     0 & 1 \\
                     -1 & 0 \\
                   \end{array}
                 \right),
~~~~T^{3}=\frac{f}{2}\left(
                   \begin{array}{cc}
                     i & 0 \\
                     0 & -i \\
                   \end{array}
                 \right).
\end{equation}

$SU(2)$ can be enlarged to $U(2)$ by appending a $U(1)$ factor and
the corresponding generator
\begin{equation}\label{iden}
T^{0}=\frac{f}{2}\left(
                   \begin{array}{cc}
                     i & 0\\
                     0 & i \\
                   \end{array}
                 \right).
\end{equation}
As a consequence,
\begin{equation}\label{Ualgebra}
[T^{0},T^{0}]=[T^{0},T^{a}]=0,~~~~[T^{a},T^{b}]=f^{ab}_{~~~c}T^{c},
\end{equation}
with $a,b,c=1,2,3$ and $f^{ab}_{~~~c}=f\epsilon^{abc}$.

Indices $0,a,b$ and so on are raised by the metric
\begin{eqnarray}
\nonumber &&h^{00}=\tr(T^{0},T^{0})=h\varpropto f^2,\\
\nonumber &&h^{0a}=h^{a0}=\tr(T^{a},T^{0})=0,\\
&&h^{ab}=\tr(T^{a},T^{b})=h\delta^{ab}\varpropto f^2\delta^{ab}.
\end{eqnarray}
In our notations, $f$ determines the normalization of structure
constants, and $h$ is the normalization constant for the metric.
Please notice that the normalization of the metric and the structure
constants should be chosen properly. Throughout this paper, for Lie
2-algebras, we normalize $h^{ab}=\delta^{ab}$ by fixing $h=1$, and
keep the normalization constant $f$ here in the structure constants.

\subsection{$SU(3)$ and $U(3)$}
The $U(3)$ algebra is described by (\ref{Ualgebra}), but with
$a,b,c=1,2,3,...8$, and the structure constants
\begin{eqnarray}\label{Ustrcon}
\nonumber &&f^{abc}=f^{[abc]},~~~~f^{123}=f,\\
\nonumber &&f^{147}=-f^{156}=f^{246}=f^{257}=f^{345}=-f^{367}=\frac{f}{2},\\
&&f^{458}=f^{678}=\frac{\sqrt{3}}{2}f.
\end{eqnarray}
In the planar matrix representation, its generators are proportional
to Gell-Mann matrices and the identity matrix,
\begin{eqnarray}\label{Gell-Mann}
\nonumber &&T^{1}=\frac{f}{2}\left(
              \begin{array}{ccc}
                0 & i & 0 \\
                i & 0 & 0 \\
                0 & 0 & 0 \\
              \end{array}
            \right),
~~~~~T^{2}=\frac{f}{2}\left(
              \begin{array}{ccc}
                0 & 1 & 0 \\
                -1 & 0 & 0 \\
                0 & 0 & 0 \\
              \end{array}
            \right),
~~~~~T^{3}=\frac{f}{2}\left(
              \begin{array}{ccc}
                i & 0 & 0 \\
                0 & -i & 0 \\
                0 & 0 & 0 \\
              \end{array}
            \right),\\
\nonumber &&T^{4}=\frac{f}{2}\left(
              \begin{array}{ccc}
                0 & 0 & i \\
                0 & 0 & 0 \\
                i & 0 & 0 \\
              \end{array}
            \right),
~~~~~T^{5}=\frac{f}{2}\left(
              \begin{array}{ccc}
                0 & 0 & 1 \\
                0 & 0 & 0 \\
                -1 & 0 & 0 \\
              \end{array}
            \right),
~~~~~T^{6}=\frac{f}{2}\left(
              \begin{array}{ccc}
                0 & 0 & 0 \\
                0 & 0 & i \\
                0 & i & 0 \\
              \end{array}
            \right),\\
\nonumber &&T^{7}=\frac{f}{2}\left(
              \begin{array}{ccc}
                0 & 0 & 0 \\
                0 & 0 & 1 \\
                0 & -1 & 0 \\
              \end{array}
            \right),
~~T^{8}=\frac{f}{2\sqrt{3}}\left(
              \begin{array}{ccc}
                i & 0 & 0 \\
                0 & i & 0 \\
                0 & 0 & -2i \\
              \end{array}
            \right),
~T^{0}=\frac{f}{\sqrt{6}}\left(
              \begin{array}{ccc}
                i & 0 & 0 \\
                0 & i & 0 \\
                0 & 0 & i \\
              \end{array}
            \right).\\
\end{eqnarray}

Getting rid of the $U(1)$ part and hence $T^{0}$, one immediately
obtains the algebra $SU(3)=U(3)/U(1)$, which is dictated by
(\ref{SUalgebra}) and (\ref{Ustrcon}).

\section{3-Algebras with a Euclidean Metric}\label{Euclidean}
In
\cite{Bagger:2006sk,Bagger:2007jr,Bagger:2007vi,Gustavsson:2007vu,Gustavsson:2008dy},
a 3-algebra
\begin{equation}\label{A4algebra1}
[T^{a},T^{b},T^{c}]=f^{abc}_{~~~~d}T^{d},~~~~(a,b,c,d=1,2,3,4)
\end{equation}
named $\mathcal{A}_4$ algebra, is constructed to describe the
dynamics of multiple M2-branes. The metric of this algebra is
positive-definite, with the signature $(+,+,+,+)$. In a suitable
basis, its metric can be chosen as $h^{ab}=\delta^{ab}$.
Interestingly, one can obtain $\mathcal{A}_4$ by lifting the $SU(2)$
algebra (\ref{SUalgebra}) with the new generator $T^{4}$. That is
\begin{eqnarray}
\nonumber [T^{4},T^{a},T^{b}]&=&f^{4ab}_{~~~~c}T^{c}=-f^{ab}_{~~~c}T^{c},\\
~[T^{a},T^{b},T^{c}]&=&f^{abc}_{~~~~4}T^{4}=f^{abc}T^{4}.~~~~(a,b,c,d=1,2,3)
\end{eqnarray}

This algebra can be easily extended by considering the
center-of-mass mode,
\begin{equation}\label{extension}
[T^{0},T^{a},T^{b}]=0.~~~~(a,b=1,2,3,4)
\end{equation}

\subsection{Lifted $SU(2)$}
As we have just shown, in the absence of the center-of-mass mode,
$\mathcal{A}_4$ can be obtained by lifting $SU(2)$ with a generator
$T^{4}$. Its structure constants are given by
\begin{equation}\label{A4strcon1}
f^{abc}_{~~~~d}=f^{[abcd]}=f\epsilon^{abcd},~~~~(a,b,c,d=1,2,3,4)
\end{equation}

Using these structure constants to solve the gauge invariance
conditions (\ref{gincond}), we find most components of the
symmetrized traces vanish, except for
\begin{equation}\label{A4SU2sol}
d^{1111}=d^{aaaa}=3d^{aabb},~~~~(a,b=1,2,3,4,~~a\neq b)
\end{equation}
and those components obtained by reordering their indices.

\subsection{Lifted $U(2)$}
If we include the center-of-mass mode and the corresponding
generator $T^{0}$, then the $SU(2)$ group is enlarged to $U(2)$. In
this way, the $\mathcal{A}_4$ algebra is also augmented as in
(\ref{extension}), resulting in a direct sum of an $\mathcal{A}_4$
with a $U(1)$ algebra. The structure constants are totally
anti-symmetric under the exchange of indices, and
\begin{eqnarray}
\nonumber &&f^{abc}_{~~~~d}=f^{abcd}=f\epsilon^{abcd},\\
&&f^{0ab}_{~~~~c}=-f^{abc}_{~~~~0}=f^{0abc}=0.~~~~(a,b,c,d=1,2,3,4)
\end{eqnarray}
Making use of them to solve (\ref{gincond}), we find the other
components of the symmetrized traces should vanish, except for the
following components
\begin{eqnarray}\label{A4U2sol}
\nonumber &&g^{000},~~~~g^{011}=g^{0aa},~~~~d^{0000},~~~~d^{0011}=d^{00aa},\\
&&d^{1111}=d^{aaaa}=3d^{aabb},~~~~(a,b=1,2,3,4,~~a\neq b)
\end{eqnarray}
and those obtained by reordering their indices, such as $g^{101}$,
$g^{110}$ etc. That is to say, the gauge invariance allows for
non-trivial solutions besides the trivial solution
(\ref{trivialsol}).

\subsection{Reduction to D2-branes}\label{A4redD2}
The above results in (\ref{A4SU2sol}) and (\ref{A4U2sol}) can be
readily understood through the gauge theory on two coincident
D2-branes. Even the undetermined components can be fixed by
reduction to D2-branes.

Taking an appropriate normalization for $\lambda_{1}$ and
$\lambda_{2}$, we can perform the ordinary trace for any planar
matrix $T$ as usual, by summing over the diagonal elements,
\begin{equation}\label{plantrace}
\tr(T)=\sum_{i}(T)_{ii}.
\end{equation}
In this way, substituting (\ref{Pauli}) and (\ref{iden}) directly
into definitions (\ref{symtrace}), one quickly get
\begin{eqnarray}\label{U2plantrace1}
\nonumber &&g^{000}=g^{011}=g^{022}=g^{033}=-\frac{if^3}{4},\\
\nonumber &&d^{0000}=d^{0011}=d^{0022}=d^{0033}=\frac{f^4}{8},\\
\nonumber &&d^{1111}=d^{2222}=d^{3333}=\frac{f^4}{8},\\
&&d^{1122}=d^{1133}=d^{2233}=\frac{f^4}{24},
\end{eqnarray}
and
\begin{equation}\label{U2plantrace2}
g^{012}=0,~~~g^{123}=0,~~~~d^{0012}=0,~~~~d^{1112}=0,~~~~d^{1123}=0,
\end{equation}
and so on.

The 3-algebras in BLG theory naturally arise from a non-associative
product, so it is difficult to imagine that their generators can be
represented by planar matrices. Indeed, Ho et. al. have emphasized
in \cite{Ho:2008bn,Ho:2007vk} that cubic matrices are a more
suitable representation for 3-algebras
\cite{Kawamura:2003cw,Kawamura:2005ic}. But if we take the cubic
matrix representation proposed in \cite{Ho:2008bn}, it seems
impossible to recover the above results.

For a single cubic matrix $T$ with three indices, the only natural
definition of a trace is the summation over diagonal elements
\cite{Kawamura:2005ic},
\begin{equation}\label{cubtrace}
\tr(T)=\sum_{i}(T)_{iii}.
\end{equation}
This definition is a trivial extrapolation of (\ref{plantrace}), and
can be trivially extrapolated to ``general matrices'' with even more
indices \cite{Kawamura:2005ic}. In \cite{Ho:2008bn} a cubic matix
representation for $\mathcal{A}_4$ generators was proposed
\begin{equation}\label{adjointrep}
(T^{a})_{ijk}=|\epsilon^{aijk}|\exp\left(\frac{i\pi}{8}\epsilon^{aijk}\right),~~~~(a,i,j,k=1,2,3,4)
\end{equation}
which meets well the algebra (\ref{A4algebra1}) under a triple
product defined as in \cite{Ho:2008bn,Kawamura:2003cw},
\begin{equation}\label{tripleprod1}
(A,B,C)_{ijk}=\sum_lA_{lij}C_{ljk}B_{lki}.
\end{equation}
But unfortunately, using these rules and this representation to
calculate (\ref{symtrace}), we simply get
\begin{equation}
g^{000}=0,~~~~g^{011}=g^{022}=g^{033}=g^{044}=0,
\end{equation}
which are at odds with (\ref{U2plantrace1}). We will return to this
point in section \ref{cubicmatrix}.

Now let us go back to see why the calculations (\ref{U2plantrace1})
and (\ref{U2plantrace2}) make sense.

Following the strategy invented in \cite{Mukhi:2008ux}, we can
perform the dimensional reduction for the Myers-Chern-Simons action
(\ref{MCS}). To get more details please refer to subsection
\ref{reductionLor} and references
\cite{Mukhi:2008ux,Gomis:2008uv,Ho:2008ei}. The resultant action
contains the required Myers-Chern-Simons terms for multiple
D2-branes and a high-dimensional term,
\begin{eqnarray}\label{MCSredD2Eucl}
\nonumber \mathcal{S}&=&\int
d^3\sigma(\mathcal{L}_{MCS}^{D2}+\mathcal{L}_{HD})\\
\nonumber \mathcal{L}_{MCS}^{D2}&=&\frac{\lambda_{1}}{g_{YM}^3}\epsilon^{\lambda\mu\nu}C_{IJK}g^{abc}D_{\lambda}X^{I}_{a}D_{\mu}X^{J}_{b}D_{\nu}X^{K}_{c}\\
\nonumber &&-\frac{\lambda_{2}}{g_{YM}^4}\epsilon^{\lambda\mu\nu}C_{8IJLMN}d^{gabc}f^{de}_{~~~g}X^{I}_{d}X^{J}_{e}D_{\lambda}X^{L}_{a}D_{\mu}X^{M}_{b}D_{\nu}X^{N}_{c},\\
\mathcal{L}_{HD}&=&-\frac{3\lambda_{1}}{g_{YM}^3}C_{8IJ}g^{abc}F^{\mu\nu}_{a}D_{\mu}X^{I}_{b}D_{\nu}X^{J}_{c},
\end{eqnarray}
with $I,J,...=1,2,3,...,7$, and $a,b,...=0,1,2,3$. We have neglected
terms involving the goldstones and $X^{I}_{\phi}$. In principle
goldstones should be eaten after a redefinition of fields. It is
remarkable that the high-dimensional term $\mathcal{L}_{HD}$ is new.
This term is hard to see from the world-volume theory of D2-branes,
but easy to obtain via dimensional reduction of M2-branes.

Because the $U(2)$ gauge group induced by D2-branes can be
represented by planar matrices (\ref{Pauli}) and (\ref{iden}), in
this situation, it is reasonable to utilize the planar matrices to
calculate the tensors $g^{abc}$ and $d^{abcd}$ with indices
$a,b,c,d=0,1,2,3$. Since the action (\ref{MCSredD2Eucl}) is reduced
from (\ref{MCS}), these components should take the same values in
them. Therefore the results (\ref{U2plantrace1}) and
(\ref{U2plantrace2}) can be trusted, even though $\mathcal{A}_4$ may
not be represented by planar matrices.

Can we also determine the values of components with the index 4?
Observing that in $\mathcal{A}_4$ algebra there is a symmetry
between $T^{1}$, $T^{2}$, $T^{3}$ and $T^{4}$, we can get these
components simply by replacing index 1 with index 4, or replacing
index 2 or 3 with index 4. This observation helps us deduce the
other components by virtue of (\ref{U2plantrace1}) and
(\ref{U2plantrace2}),
\begin{equation}
g^{044}=-\frac{if^3}{4},~~~~d^{0044}=\frac{f^4}{8},~~~~d^{4444}=\frac{f^4}{8},~~~~d^{1144}=\frac{f^4}{24},~~~~...
\end{equation}
Hence we have proved solution (\ref{A4U2sol}) and fixed the
non-vanishing components.

\section{3-Algebras with a Lorentzian Metric}\label{Lorentzian}
Although $\mathcal{A}_4$ successfully incorporates the $SU(2)$
algebra, its generalization to $SU(N)$ for an arbitrary $N$ turns
out to be very difficult. If one requires
\begin{enumerate}
\item there is an invariant positive-definite metric,
\item the structure constants are totally anti-symmetric and satisfy the fundamental
identity,
\item the 3-algebra is finite, non-trivial and irreducible,
\end{enumerate}
then $\mathcal{A}_4$ algebra is the unique 3-algebra
\cite{Gustavsson:2008dy,Bandres:2008vf,Ho:2008bn,Papadopoulos:2008sk,Gauntlett:2008uf}.
Such a ``uniqueness theorem'' can be bypassed in many ways via
relaxing some of the above requirements. Very recently, permitting
at least one negative signature in the metric, three groups
\cite{Gomis:2008uv,Benvenuti:2008bt,Ho:2008ei} independently
introduced another way to lift any Lie 2-algebra $\mathcal{G}$ to a
3-algebra.

For arbitrary Lie 2-algebra
\begin{equation}
[T^{a},T^{b}]=f^{ab}_{~~~c}T^{c},
\end{equation}
by introducing two new generators $T^{+}$ and $T^{-}$, they lifted
it to a 3-algebra \cite{Gomis:2008uv,Benvenuti:2008bt,Ho:2008ei}
\begin{eqnarray}\label{Loralgebra}
\nonumber [T^{-},T^{a},T^{b}]&=&0,\\
\nonumber [T^{+},T^{a},T^{b}]&=&f^{+ab}_{~~~~c}T^{c}=f^{ab}_{~~~c}T^{c},\\
~[T^{a},T^{b},T^{c}]&=&f^{abc}_{~~~~-}T^{-}=f^{abc}T^{-}.
\end{eqnarray}

When $\mathcal{G}$ is positive-definite, the metric of such a
3-algebra is not positive-definite, but has a Lorentzian signature
$(-,+,+,...+)$. Being interested in Myers-Chern-Simons action for this
class of algebras, we will take $\mathcal{G}$ to be $SU(N)$ or
$U(N)$ and study some concrete examples. In a suitable basis, the
metric of a lifted $U(N)$ algebra is of the form
\cite{Gomis:2008uv,Benvenuti:2008bt,Ho:2008ei}
\begin{equation}\label{Lormetric}
h^{+-}=-1,~~~~h^{\pm\pm}=0,~~~~h^{ab}=\delta^{ab},~~~~h^{a\pm}=0,~~~~(a,b=0,1,2,...N^2).
\end{equation}

Throughout this section, we always assume the basis is defined such
that the metric assumes the form in (\ref{Lormetric}).

\subsection{Lifted $SU(2)$}\label{lorSU(3)}
Using the structure constants of $SU(2)$ together with
(\ref{Loralgebra}) and (\ref{Lormetric}), one can get the
corresponding structure constants with four indices. Solving the
gauge invariance conditions (\ref{gincond}), we find most components
of the symmetrized traces should vanish, except for
\begin{eqnarray}
\nonumber &&g^{aa+}=-\frac{1}{2}g^{-++},~~~~g^{+++},~~~~d^{++++},~~~~d^{aa++}=-\frac{1}{3}d^{-+++},\\
&&d^{aabb}=-d^{aa-+}=\frac{1}{3}d^{aaaa}=\frac{1}{2}d^{--++},~~~~(a,b=1,2,3,~~a\neq b)
\end{eqnarray}
and those components obtained by reordering their indices.

\subsection{Lifted $U(2)$}
Following a similar procedure, we get the solutions
to (\ref{gincond}). Except for the components
\begin{eqnarray}
\nonumber &&g^{000},~~~~g^{00+},~~~~g^{0aa}=-g^{0-+},~~~~g^{0++},~~~~g^{aa+}=-\frac{1}{2}g^{-++},~~~~g^{+++},\\
\nonumber &&d^{0000},~~~~d^{000+},~~~~d^{00++},~~~~d^{0+++},~~~~d^{++++},\\
\nonumber &&d^{00aa}=-d^{00-+},~~~~d^{0aa+}=-\frac{1}{2}d^{0-++},~~~~d^{aa++}=-\frac{1}{3}d^{-+++},\\
&&d^{aabb}=-d^{aa-+}=\frac{1}{3}d^{aaaa}=\frac{1}{2}d^{--++},~~~~(a,b=1,2,3,~~a\neq b)
\end{eqnarray}
and those obtained by reordering their indices, the other components
of the symmetric tensors (\ref{symtrace}) must vanish in order to
respect the gauge invariance.

In subsection \ref{lorSU(3)} and here, some of the components can be
also be fixed as before,
\begin{eqnarray}
\nonumber &&g^{000}=g^{0aa}=-\frac{if^3}{4},~~~~d^{0000}=d^{00aa}=\frac{f^4}{8},\\
\nonumber &&d^{aaaa}=\frac{f^4}{8},~~~~d^{aabb}=\frac{f^4}{24}.~~~~(a,b=1,2,3,~~a\neq b)
\end{eqnarray}

\subsection{Lifted $SU(3)$}
In this example, the structure constants of the 3-algebra can be
obtained by combining (\ref{Ustrcon}), (\ref{Loralgebra}) and
(\ref{Lormetric}). Solving the gauge invariance conditions
(\ref{gincond}), it turns out that the surviving components of the
symmetrized traces are
\begin{eqnarray}
\nonumber &&g^{aa+}=-\frac{1}{2}g^{-++},~~~~g^{+++},~~~~d^{++++},~~~~d^{aa++}=-\frac{1}{3}d^{-+++},\\
&&d^{aabb}=-d^{aa-+}=\frac{1}{3}d^{aaaa}=\frac{1}{2}d^{--++},~~~~(a,b=1,2,3,...,8,~~a\neq b)
\end{eqnarray}
and those obtained by reordering their indices.

\subsection{Lifted $U(3)$}
Starting with $U(3)$ algebra, repeating the above procedure, one
finds the other components of (\ref{symtrace}) should vanish, except
for
\begin{eqnarray}
\nonumber &&g^{000},~~~~g^{00+},~~~~g^{0aa}=-g^{0-+},~~~~g^{0++},~~~~g^{aa+}=-\frac{1}{2}g^{-++},~~~~g^{+++},\\
\nonumber &&d^{0000},~~~~d^{000+},~~~~d^{00++},~~~~d^{0+++},~~~~d^{++++},\\
\nonumber &&d^{00aa}=-d^{00-+},~~~~d^{0aa+}=-\frac{1}{2}d^{0-++},~~~~d^{aa++}=-\frac{1}{3}d^{-+++},\\
&&d^{aabb}=-d^{aa-+}=\frac{1}{3}d^{aaaa}=\frac{1}{2}d^{--++},~~~~(a,b=1,2,3,...,8,~~a\neq b)
\end{eqnarray}
and those obtained by reordering their indices.

Thanks to Gell-Mann matrices (\ref{Gell-Mann}), we can fixed some
components as
\begin{eqnarray}
\nonumber &&g^{000}=g^{0aa}=-\frac{if^3}{2\sqrt{6}},~~~~d^{0000}=d^{00aa}=\frac{f^4}{12},\\
\nonumber &&d^{aaaa}=\frac{f^4}{8},~~~~d^{aabb}=\frac{f^4}{24}.~~~~(a,b=1,2,3,~~a\neq b)
\end{eqnarray}

\subsection{Lifted $SU(N)$ and Lifted $U(N)$}
The results of the above examples are very suggestive for a
generalization. They suggest that for an $SU(N)$-lifted 3-algebra
with a Lorentzian metric, in order to ensure the gauge invariance
(\ref{gincond}), the non-vanishing components of symmetrized traces
can only be
\begin{eqnarray}
\nonumber &&g^{aa+}=-\frac{1}{2}g^{-++},~~~~g^{+++},~~~~d^{++++},~~~~d^{aa++}=-\frac{1}{3}d^{-+++},\\
&&d^{aabb}=-d^{aa-+}=\frac{1}{3}d^{aaaa}=\frac{1}{2}d^{--++},~~~~(a,b=1,2,3,...,N^2-1,~~a\neq b)
\end{eqnarray}
and those obtained by reordering their indices. For the 3-algebra
extended from $U(N)$, the surviving components are
\begin{eqnarray}
\nonumber &&g^{000},~~~~g^{00+},~~~~g^{0aa}=-g^{0-+},~~~~g^{0++},~~~~g^{aa+}=-\frac{1}{2}g^{-++},~~~~g^{+++},\\
\nonumber &&d^{0000},~~~~d^{000+},~~~~d^{00++},~~~~d^{0+++},~~~~d^{++++},\\
\nonumber &&d^{00aa}=-d^{00-+},~~~~d^{0aa+}=-\frac{1}{2}d^{0-++},~~~~d^{aa++}=-\frac{1}{3}d^{-+++},\\
&&d^{aabb}=-d^{aa-+}=\frac{1}{3}d^{aaaa}=\frac{1}{2}d^{--++},~~~~(a,b=1,2,3,...,N^2-1,~~a\neq b)
\end{eqnarray}
and those obtained by reordering their indices.

We stress the basis of generators we employed here. When writing
down the above expressions, we have assumed the basis is defined
such that the metric takes the form in (\ref{Lormetric}).

On the one hand, we have proved these results with $N=2,3$. On the
other hand, a large class of gauge groups can be restricted to a
subalgebra generated by $SU(2)$ or $SU(3)$. Hence the gauge
invariance conditions presented in this subsection is not hard to
understand.

Although we cannot prove the full constraints (\ref{gincond}) be
satisfied, we have checked some relations. To show a few of them, we
rewrite the first equation of (\ref{gincond}) as
\begin{equation}\label{checkUN1}
g^{D(AB}f^{C)FE}_{~~~~~~D}=0,~~~~(A,B,...=+,-,0,1,2,3,...,N^2-1)
\end{equation}
Suppose for certain $a,e,f$, we have $f^{fe}_{~~~d}\neq0$ if and
only if $d=a$. This is the case when we work in a basis such as the
Chevalley basis. Then the following components of (\ref{checkUN1})
lead to
\begin{eqnarray}
\nonumber (ABCEF)=(a00ef)&\Rightarrow&g^{00-}=0,\\
\nonumber (ABCEF)=(a0+ef)&\Rightarrow&g^{0aa}+g^{0-+}=0,\\
\nonumber (ABCEF)=(a++ef)&\Rightarrow&2g^{aa+}+g^{-++}=0,\\
\nonumber (ABCEF)=(e+++f)&\Rightarrow&g^{a++}=0,\\
...&&...
\end{eqnarray}
In this way, much more components can be write down. We think the
strategy depicted above and the relation between $SU(2)$ and $SU(N)$
are the key points to get a refined proof.

\subsection{Reduction to D2-branes}\label{reductionLor}
The reduction of M2-branes to D2-branes is parallel to the previous
section. The strategy is to choose a vacuum
\cite{Mukhi:2008ux,Gomis:2008uv,Ho:2008ei}
\begin{equation}
<X^{-8}>=g_{YM}
\end{equation}
with other scalar fields to be zero, and define
\begin{equation}\label{redGP}
A_{\mu}^{a-}=\frac{1}{2}A_{\mu}^{a},~~~~\frac{1}{2}\epsilon^{a}_{~bc}A_{\mu}^{bc}=B_{\mu}^{a}.~~~~(a,b,c\neq-)
\end{equation}
Using the equation of equation of $B_{\mu}^{a}$ (see
\cite{Mukhi:2008ux,Gomis:2008uv,Ho:2008ei}) at the leading order of
$g_{YM}^{-1}$, and then rescaling $X\rightarrow X/g_{YM}$, one can
get the reduced action. Since $B_{\mu}^{a}$ couples in
Myers-Chern-Simons action (\ref{MCS}) in the form of higher order,
its equation of motion at the leading order is unchanged\footnote{To
make the convention of notations in accordance with
(\ref{MCSredD2Lor}), when writing down (\ref{redGP}) and
(\ref{eomB}) we have already rescaled $A\rightarrow A/2$ and
$X\rightarrow X/g_{YM}$ as did in \cite{Mukhi:2008ux}.}
\cite{Mukhi:2008ux,Gomis:2008uv,Ho:2008ei},
\begin{equation}\label{eomB}
B_{\mu}^{a}=\frac{1}{4g_{YM}^2}\epsilon_{\mu}^{~\nu\lambda}F^{a}_{\nu\lambda}-\frac{1}{2g_{YM}^2}D_{\mu}X^{8}_{a}.
\end{equation}

Neglecting terms related to goldstones and $X^{I}_{-}$, we write down the reduced
action as follows
\begin{eqnarray}\label{MCSredD2Lor}
\nonumber \mathcal{S}&=&\int d^3\sigma(\mathcal{L}_{MCS}^{D2}+\mathcal{L}_{HD}),\\
\nonumber \mathcal{L}_{MCS}^{D2}&=&\frac{\lambda_{1}}{g_{YM}^3}\epsilon^{\lambda\mu\nu}C_{IJK}g^{abc}D_{\lambda}X^{I}_{a}D_{\mu}X^{J}_{b}D_{\nu}X^{K}_{c}\\
\nonumber &&+\frac{\lambda_{2}}{g_{YM}^4}\epsilon^{\lambda\mu\nu}C_{8IJLMN}d^{gabc}f^{de}_{~~~g}X^{I}_{d}X^{J}_{e}D_{\lambda}X^{L}_{a}D_{\mu}X^{M}_{b}D_{\nu}X^{N}_{c},\\
\mathcal{L}_{HD}&=&-\frac{3\lambda_{1}}{g_{YM}^3}C_{8IJ}g^{abc}F^{\mu\nu}_{a}D_{\mu}X^{I}_{b}D_{\nu}X^{J}_{c},
\end{eqnarray}
with $I,J,...=1,2,3,...,7$ and $a,b,c...=0,1,2,...,N^2-1$. Higher
order terms with respect to $g_{YM}^{-1}$ are neglected because the
reduction is done in the limit $g_{YM}\rightarrow\infty$. We also
removed the ghost fields by setting $X^{I}_{+}=constant$
\cite{Ho:2008ei}.

Notice here the second term of $\mathcal{L}_{MCS}^{D2}$ takes a
different sign from that in (\ref{MCSredD2Eucl}). This is because
structure constants $f^{4de}_{~~~~g}$ in $\mathcal{A}_4$ and
$f^{+de}_{~~~~g}$ in the present 3-algebras relate to
$f^{de}_{~~~g}$ of Lie 2-algebras in different ways,
\begin{equation}
f^{4de}_{~~~~g}=-f^{de}_{~~~g},~~~~f^{+de}_{~~~~g}=f^{de}_{~~~g}.
\end{equation}

The surviving terms contain Myers-Chern-Simons terms for D2-branes.
These are exactly what we expected. Once again we get a
high-dimensional term $\mathcal{L}_{HD}$. This term is proportional
to $g_{YM}^{-3}$, which is of the leading order in
(\ref{MCSredD2Lor}).

\section{On Cubic Matrices}\label{cubicmatrix}
In subsection \ref{A4redD2}, we saw that under a simple definition
of trace (\ref{cubtrace}), the cubic matrix representation
(\ref{adjointrep}) fails to reproduce the required non-vanishing
components. Some comments are needed here.

First, one can make use of the representation (\ref{adjointrep}) and
rule (\ref{tripleprod1}) to work out the symmetrized triple product
\begin{equation}
\{T^{a},T^{b},T^{c}\}_{ijk}=6(T^{(a},T^{b},T^{c)})_{ijk}=-i\sum_{d}|\epsilon^{abcd}|\epsilon^{dijk}(T^{d})_{ijk}.
\end{equation}
This expression and the results in section \ref{Euclidean} would be
helpful for exploring a possibly better definition of trace instead
of (\ref{cubtrace}).

Second, the definition (\ref{tripleprod1}) of the triple product is
not unique in principle. Here are two alternative schemes:
\begin{equation}\label{tripleprod2}
(A,B,C)_{lmn}=\sum_{i,j,k}A_{lij}B_{mjk}C_{nki},
\end{equation}
\begin{equation}\label{tripleprod3}
(A,B,C)_{lmn}=\sum_{i,j}A_{lij}B_{mij}C_{nij},
\end{equation}
which are easy to be generalized for multiple products of cubic
matrices. Their drawback is: if we take the structure constants in
(\ref{A4algebra1}) as a generalized ``adjoint
representation''\footnote{The cubic matrices (\ref{adjointrep}) play
a similar role.} of $\mathcal{A}_4$ algebra, then only the
anti-symmetrized product based on (\ref{tripleprod1}) can recover
$\mathcal{A}_4$ algebraic relation (\ref{A4algebra1}). But they also
have their virtues. Suppose a certain generator $T^{a}$ in the cubic
representation satisfying
\begin{equation}\label{nullcond}
(T^{a})_{ijk}=(T^{a})_{jki}=(T^{a})_{kij},~~~~(T^{a})_{ijj}=0.
\end{equation}
When (\ref{tripleprod1}) is applied, this will inevitably lead to
$g^{0aa}=0$, which is at odds with (\ref{U2plantrace1}). In
contrast, neither (\ref{tripleprod2}) nor (\ref{tripleprod3})
suffers for such a problem.

The above problem also inspires us to search for cubic matrix
representations $(T^{a})_{ijj}\neq0$ breaking the condition
(\ref{nullcond}). This is the third possibility one may try to
translate the 3-algebraic rules into the language of cubic matrices.

\section{Conclusion}\label{conclusion}
In this paper, we studied Myers-Chern-Simons action for
multiple M2-branes. We wrote down the gauge invariance
conditions and solved them in some concrete examples. One example is
the gauge theory based on $\mathcal{A}_4$ algebra. The other class
of examples are non-Abelian gauge theories by lifting $SU(N)$ or
$U(N)$ to 3-algebras with a Lorentzian signature. The reduction of
M2-branes to D2-branes puts addtional constraints on the
Myers-Chern-Simons action.

We studied simpler examples first and checked a few relations for the
lifted $U(N)$, In the case of $\mathcal{A}_4$, we found all of the components can be fixed up to
an overall normalization. For 3-algebras with a Lorentzian
signature, only some components can be fixed.

We also offered some tentative thoughts on cubic matrices as the
representations of 3-algebras. In spite of notorious difficulty,
this is a direction deserving to follow up in the future. We plan to study some
physical consequences of the terms introduced in this paper, and their
supersymmetric generalization in a future paper.

{\bf Acknowledgement}: We would like to thank Yushu Song for
substantial discussions and Qin-Yan Tan for kind help in programme.
This work was supported by grants of CNSF and grants of USTC.

\end{document}